\documentstyle[epsf]{mn1}

\title[$\Delta T_{\rm SZ}$ vs $L$ Relation]{An SZ Temperature Decrement - X-ray Luminosity
Relation for Galaxy Clusters.}

\author[Cooray]{Asantha ~R. ~Cooray\\
Department of Astronomy and Astrophysics, University of Chicago, Chicago IL 60637. E-mail: asante@hyde.uchicago.edu}

\date{Submitted: 12 March 1998; Accepted: 22 March 1999}
\begin{document}
\maketitle

\begin{abstract}
We present the observed relation between
$\Delta T_{\rm SZ}$, the cosmic microwave
background (CMB) temperature decrement due to the 
Sunyaev-Zeldovich (SZ) effect,
and $L$, the X-ray luminosity of galaxy clusters.
We discuss this relation in terms of the 
cluster properties, and
show that the slope of the observed $\Delta T_{\rm SZ}-L$ 
relation is in agreement
with both the $L-T_{\rm e}$ relation based on numerical simulations
and X-ray emission observations, and the $M_{\rm gas}-L$ relation
based on observation. 
The slope of the $\Delta T_{\rm SZ}-L$ relation is also consistent
with the $M_{\rm tot}-L$ relation, where $M_{\rm tot}$ is the cluster
total mass based on gravitational lensing observations. 
This agreement may be taken to imply a constant gas mass fraction
within galaxy clusters, however, there are large uncertainties,
dominated by observational errors, associated with these
relations. Using the $\Delta T_{\rm SZ}-L$ relation and the cluster
X-ray luminosity function, we evaluate the local cluster contribution to
arcminute scale cosmic microwave background anisotropies. 
The Compton distortion 
$y$-parameter produced by galaxy clusters through SZ effect is
roughly two orders of magnitude lower than the current upper limit based
on {\it FIRAS} observations.

\end{abstract}
%------------------------------------------------------------------------------
% User-supplied List of keywords.


%------------------------------------------------------------------------------
\section{Introduction}

The X-ray emission from galaxy clusters has enabled the study
of what is assumed to be the largest virialized systems in the universe. 
The X-ray electron temperature, $T_{\rm e}$,
 measures the depth of the galaxy cluster potential, while
the X-ray luminosity, $L$, emitted as thermal bremsstrahlung by the
intracluster plasma, measures primarily the baryonic number density within
this potential. When combined with the fact that the
electron temperature is a robust estimator of the galaxy cluster total mass,
$M_{\rm tot}$, the X-ray emission observations can be used
in cosmological studies to derive constraints on the power spectrum of the
initial density perturbations and on cosmological parameters
(e.g., Eke {\it et al.} 1998; Oukbir \& Blanchard 1997; Bahcall,
Fan \& Cen 1997). However, most of the important conclusions from
such studies are dependent on the accuracy of used scaling
relations between the cluster properties, such as the cluster X-ray
luminosity, total mass, gas mass,
and the electron temperature. Independent of the large scale distribution, 
baryonic component of the individual
clusters has been used to constrain the mass density
of the universe, $\Omega_m$ (e.g., Briel et al. 1992; 
White et al. 1993;  White \& Fabian 1995; David,
Jones \& Forman 1995; Evrard 1997), but
such studies can be subject to variations in physical
properties from one cluster to another, e.g., due to cluster
cooling flows (Fabian et al. 1994; White, Jones
\& Forman 1997; Allen \& Fabian 1998; Markevitch 1998).

Apart from X-ray emission observations, 
another well known probe of the intracluster gas
distribution is the Sunyaev-Zeldovich (SZ) effect (Sunyaev \&
Zeldovich 1970, 1980). The SZ effect is the scattering of 
the cosmic microwave background (CMB) radiation
via hot electrons of the X-ray emitting gas through an inverse-Compton
process, producing
a distortion in the CMB spectrum. 
In recent years,
increasingly sensitive observations of galaxy clusters, first with
single-dish telescopes and now with interferometers, have
produced accurate maps of the CMB temperature
change resulting from the SZ effect.
Unlike the X-ray emission from galaxy clusters,  
the magnitude of the CMB temperature decrement due to
the SZ effect is independent of the redshift, and
allows a direct probe to the distant universe. 
It is likely that the potential use of the SZ effect as a cosmological probe to distant universe
is yet to be fully realized, and more
work, both together and independent of the X-ray properties
of galaxy clusters, is required. The combination of X-ray emission and SZ
effect towards a given cluster can be used to determine the distance
to that cluster, from which the Hubble constant can be derived. Also,
the SZ effect, which measures that gas mass within galaxy clusters, can be
used to constrain the total mass density of the universe
(see, e.g., Myers et al. 1997).
Other than cosmological uses, the SZ effect and the X-ray emission from
clusters can be used to study cluster gas physics, since these two
observables depends differently on the 
electron number density and 
temperature distributions.

Given the ever increasing galaxy cluster SZ data, 
we have initiated a study to investigate the  ways in which
the SZ effect can be used both as a cosmological tool, as well
as a tool to
understand the distribution and physical properties 
of baryonic content within clusters. Here in this paper,
we present the observed relation between
the CMB temperature change due to the SZ effect, $\Delta T_{\rm SZ}$, 
and the X-ray luminosity, $L$, of a sample of galaxy clusters. In Section 2,
we present a brief introduction to the SZ effect and the 
X-ray emission, and
formulate the expected relation between $\Delta T_{\rm SZ}$ and $L$.
In Section 3, we present the used cluster sample in which SZ effect has been
observed and derive the $\Delta T_{\rm SZ} - L$ relation, while  
in Section 4,
we discuss the observed relation in terms of
the galaxy cluster properties. In the same section, we 
use this relation as an application to study the cluster contribution
to arcminute scale
cosmic microwave background anisotropies.

\section{Theory}

\setcounter{table}{0}
\begin{table*}
\begin{minipage}{125mm}
\caption{SZ Cluster Sample}
\begin{tabular}{@{}lccccl}
\hline
Cluster Name   & z &
${\rm L}$ & 
${\rm T_{\rm e}}$ &
${\rm \Delta T^{\rm RJ}_{\rm SZ}}$  &
Reference \\
   &  &
$({\rm h^{-2}_{\rm 50}\, 10^{44}\, ergs\, s^{-1}})$ &
(keV) &
(mK) & \\
\hline
Coma (A1656) & 0.0232 & 11.3 & 8.2 $\pm$ 0.1 & -0.55 $\pm$ 0.10 & 1,1,2 \& 3 \\
A2256 & 0.0601 & 11.1 & 7.5 $\pm$ 0.2 & -0.44 $\pm$ 0.09 & 1,1,2\\
A2142 & 0.0899 & 30.4 & 9.3 $\pm$ 1.0 & -0.90 $\pm$ 0.14 & 4,4,2\\
A478 & 0.09 & 24.2 & 8.1 $\pm$ 1.0 & -0.92 $\pm$ 0.15 & 4,4,2\\
A1413 & 0.143 & 15.4 & 8.5 $\pm$ 1.0 & -0.96 $\pm$ 0.11 & 4,4,5  \\
A2204 & 0.152 & 34.6 & 9.2 $\pm$ 2.5 & -0.96 $\pm$ 0.28 & 4,4,15\\
A2218 & 0.171 & 10.8 & 7.05 $\pm$ 0.35 & -0.75 $\pm$ 0.20 & 4,4,6\\
      &       &      &                        & -0.90 $\pm$ 0.10 & 7 \\
A1689 & 0.180 &  32.2 & 10.0 $\pm$ 1.0 & -1.91 $\pm$ 0.30 & 4,4,8\\
A665 & 0.182 & 17.8 & 9.03 $\pm$ 0.60 &  -0.91 $\pm$ 0.09& 4,4,9\\
A773 & 0.197 & 14.8 & 9.29 $\pm$ 0.65 & -0.89 $\pm$ 0.10 & 4,4,10\\
     &       &      &                      & -0.67 $\pm$ 0.10 & 11\\
A2163 & 0.201 &  60.0  & 13.8 $\pm$ 0.8 & -1.93 $\pm$ 0.28 & 4,4,8\\
A1835 & 0.252 &  44.9 &  9.8 $\pm$ 2.5  & -1.34 $\pm$ 0.15 & 4,4,15\\
Z3146 & 0.291 &  37.3 &  11.3 $\pm$ 3.5 & -0.86 $\pm$ 0.14 & 4,4,15\\
Cl0016 & 0.541 & 26.2 & 7.55 $\pm$ 0.7  & -1.21 $\pm$ 0.19 & 12 \& 13,14,14\\
       &      &       &              & -0.75 $\pm$ 0.08  & 11\\
\hline
\end{tabular}
\smallskip

Notes: column (6) indicates references for the X-ray luminosity, temperature 
and SZ measurement, respectively (for clusters with multiple SZ measurements,
the second column under the same cluster refers only to the SZ data):
1. Arnaud \& Evrard (1998);
2. Myers et al. (1997);
3. Herbig et al. (1995);
4. Allen \& Fabian (1998);
5. Grainge et al. (1996);
6. Uyaniker  et al. (1997); 
7. Jones (1995);
8. Holzapfel et al. (1997); 
9. Birkinshaw  et al. (1991);
10. Grainge et al. (1993);
11. Carlstrom et al. (1996);
12. Neumann \& B\"ohringer (1997);
13. Tsuru et al. (1996);
14. Hughes \& Birkinshaw (1998);
15. Holzapfel (1996);
\end{minipage}
\end{table*}

Briefly, the SZ effect is
a distortion of the cosmic microwave background
(CMB) radiation by inverse-Compton scattering of
thermal electrons within the hot intracluster medium
(Sunyaev \& Zeldovich 1970, 1980).
The change in the CMB brightness temperature observed
is:
\begin{equation}
\frac{\Delta T}{T_{\rm CMB}} = f(x)
\int \left(\frac{k_B T_e}{m_e c^2}\right) n_e \sigma_T dl,
\end{equation}
where
\begin{equation}
f(x) = \left[ \frac{x (e^{x}+1)}{e^{x}-1} -4 \right]
\end{equation}
is the frequency dependence with
$x = h \nu/k_B T_{\rm CMB}$, $T_{\rm CMB} = 2.728 \pm 0.002$
(Fixsen {\it et al.} 1996) and $n_e$, $T_e$ and $\sigma_T$ are the       
electron density, electron temperature and cross section for the Thomson
scattering. The integral is performed along the line of sight through the
cluster. We refer the reader to
a recent review by Birkinshaw (1998) on the SZ effect and its observation.
At the Rayleigh-Jeans (RJ) part of the frequency spectrum, $f(x)=-2$.

The other important
observable of the hot intracluster gas is the 
X-ray emission, whose surface brightness $S_X$ can be written
as:
\begin{equation}
S_X = \frac{1}{4 \pi (1+z)^3} \int n ^{2}_{e} \Lambda_e dl,
\end{equation}
where $z$ is the redshift and $\Lambda_e(\Delta E,T_e)$ is the
X-ray spectral emissivity of the cluster gas due to thermal
Bremsstrahlung emission within a certain energy band $\Delta E$ 
($\Lambda_e \propto T_{e}^{1/2}$).
By combining the SZ intensity change and the X-ray emission observations
of a given cluster,
the angular diameter 
distance to the cluster can be derived due to the          
different dependence of the X-ray emission and the SZ effect
on the electron number density, $n_e$
(e.g., Cavaliere {\it et al.} 1977).
Combining the distance measurement with redshift
allows a determination of the
Hubble constant, H$_0$, as a function of certain cosmological
parameters (e.g., see, Hughes \& Birkinshaw 1998).  
Other than the derivation of the Hubble constant, SZ effect and
X-ray emission can
in principle be used to constrain the cosmological parameters
based on distance measurements of
a large sample of galaxy clusters with a wide range of redshift.

The present paper discusses the relation between $\Delta T_{\rm SZ}$
and $L$, the X-ray luminosity, for a sample of galaxy clusters.
We can estimate the expected relation based on
$n_{e}$ and $T_{e}$ dependences between the SZ effect and the
X-ray emission.
The SZ effect is due to the pressure
integrated along the line of sight, i.e. $\Delta T_{\rm  SZ} \propto
n_{\rm e} T_{\rm e}$, while the X-ray emission is
 due to the thermal Bremsstrahlung with
$L \propto n_{\rm e}^2 T_{\rm e}^{1/2}$. 
Here, we igonore the contribution from X-ray line emission. By removing the
$n_{\rm e}$ dependence between $\Delta T_{\rm SZ}$ and $L$,
one gets $\Delta T_{\rm SZ} \propto L^{1/2} T_{\rm e}^{3/4}$.
The relation between $L$ and $T_{\rm e}$ has been well studied
based on numerical simulations (e.g., Cavaliere {\it et al.} 1997)
and observed data (e.g., Mushotzky \&
Scharf 1997; Allen \& Fabian 1998; Arnaud \& Evrard 1998). Assuming that
$L \propto T_{\rm e}^{\alpha}$,
\begin{equation}
\Delta T_{\rm SZ} \propto L^{\frac{1}{2}+\frac{3}{4 \alpha} }
\end{equation}
Based on numerical simulations, Cavaliere {\it et al.}  (1997)
predicts that $\alpha=5$ at the scale of groups and flattens to
$\alpha=3$ for rich clusters, and saturates to
$\alpha=2$ for high temperature clusters. Currently,
the SZ effect has been detected towards rich
clusters with moderate to high electron temperatures, 
resulting in $\Delta T_{\rm SZ} \propto L^{0.65}$ to 
$L^{0.88}$, when $\alpha$ varies from 5 to 2.
The observed data currently suggest that $\alpha \sim 2.8$, 
less than previous studies which suggested values for $\alpha \sim 3$
to 3.3 (e.g., David et al. 1993). 

Also, since $\Delta T_{\rm SZ}$ is a measurement of the pressure along the line
of sight to the cluster, the integral of $\Delta T_{\rm SZ}$ 
through a cylindrical cut of the cluster is directly proportional to the gas mass within that
cylinder. This assumes that the cluster gas is isobaric. If we expect
the gas mass to scale with luminosity according to a relation of the
form $M_{\rm gas} \propto L^{\gamma}$, then,
\begin{equation}
\Delta T_{\rm SZ} \propto L^{\gamma}.
\end{equation}
The second estimate for the $\Delta T_{\rm SZ}-L$
relation should be more accurate since the
measurement of gas mass accounts for the integrated flux along the
line of sight, while the former estimate relies slightly on the assumption of
spherical geometry for galaxy clusters and that no cluster gas
clumping is present. 

\section{Data}

In order to derive the observed relation
between $\Delta T_{\rm SZ}$ and $L$, we have compiled a sample of 
galaxy clusters for which SZ data are available from literature, and
for which accurate X-ray results are also available. We list these
clusters, X-ray luminosities 
in the 2 to 10 keV band, electron temperature, 
SZ temperature decrement and references to each of the
X-ray luminosity, temperature and SZ experiments in Table 1.

Given that the tabulated SZ observations were made at different
frequencies, we have scaled the
SZ temperature decrement to the RJ part of the spectrum so that 
the comparison between the SZ effect and the X-ray luminosity of 
galaxy clusters is uniform.
Also, some of the SZ experiments ignored the higher order corrections to
the inverse-Compton scattering in evaluating the SZ temperature decrement.
Using the relativistic corrections presented by Itoh {\it et al.}
(1997), we have corrected for the SZ temperature decrement, which are
mainly important for clusters with high electron temperatures ($>$ 10 keV). 
Such corrections only change the SZ temperature
decrement by about 3\% to 5\%, but the 
relation between $\Delta T_{\rm SZ}$ and $L$  presented here
is not heavily dependent on such corrections.

Another uncertainty associated with the tabulated values is the
accuracy of X-ray luminosities. Except for Cl0016, we have used the published
luminosity values from Allen \& Fabian (1998) and Arnaud \& Evrard (1998).
These studies have taken into account the variations in the X-ray
luminosities due to cluster cooling flows, so the values tabulated here should
be unbiased from cooling flow effects. However, we note that these
luminosities may have both statistical and systematic uncertainties of
the order 15\%, resulting from poor calibration to
uncertainties associated with removal of individual cluster
cooling flow regions in calculating the luminosities.
Our SZ cluster sample is primarily made of interferometric observations of the
SZ effect (Carlstrom et al. 1996; Grainge et al. 1993), SuZIE
observations at the Caltech Sub-mm Observatory (Holzapfel et al. 1997), and
OVRO single-dish observations of the low redshift clusters (Myers et
al. 1997). 

The published SZ temperature decrement values in Myers et al.
(1997) are not corrected for the beam dilution and switching. However, based
 on modeling of the cluster gas distribution for individual clusters, the
authors have calculated various efficiencies in this process (see, Table 8 in Myers
et al. 1997). We used these values to convert the observed SZ
temperature decrements to beam-corrected values, which are presented in
Table 1.

\begin{figure}
\vspace{5cm}
\caption{The observed $\Delta T_{\rm SZ}$ vs. $L$,
with the best fit relation shown as
a solid line (Eq.\ 7). The open circles
represent the low redshift clusters from Myers et al. (1997), while the filled 
cirlces represent the clusters between $0.1 < z < 0.21$. The solid triangle
is Cl0016.}
\end{figure}

In Fig.\ 1, we show the observed SZ temperature decrement
vs. the X-ray luminosity in the 2 to 10 keV band, to which we have fitted
a relation of the form $\Delta T_{\rm SZ} = A L^{\kappa}$. 
For clusters with multiple SZ measurements, we have used the mean value
of the reported temperature decrement but have appropriately scaled the
uncertainty so that it covers the whole range suggested by the two separate
measurements. Using a maximum-likelihood minimization, the data are best 
explained by:
\begin{equation}
\Delta T_{\rm SZ} = -(0.17 \pm 0.10) \left(\frac{L}{10^{44}\,
h_{50}^{-2}\, {\rm ergs\, s^{-1}}}\right)^{0.61 \pm 0.18}\, {\rm mk},
\end{equation}
where the uncertainty is the 1$\sigma$ statistical error.

\begin{figure}
\vspace{5cm}
\caption{The observed $M_{\rm gas}$ vs. $L$, with the 
best fit relation shown as a solid line (Eq.\ 8).} 
\end{figure}

In order to test the expected $\Delta T_{\rm SZ}$ and $L$ relation based
on the X-ray emission data through measurements of the cluster gas mass,
we complied a list of gas mass estimates from published data.
In Fig.\ 2, we show the gas mass within 0.5 Mpc of the cluster center as
derived by White et al. (1997) for a sample of $\sim$ 150 clusters. 
The derived relation between gas mass, $M_{\rm gas}$, and
the cluster luminosity, $L$, is:
\begin{eqnarray}
M_{\rm gas} = (21.42 \pm 3.16) &\left( \frac{L}{10^{44}\, h_{50}^{-2}\,
{\rm ergs\, s^{-1}}}\right)^{0.66 \pm 0.06}\, \\ \nonumber
& \times 10^{12} {\rm h_{50}^{-2.5}}\, {\rm M_{\sun}},
\end{eqnarray}
which is very similar to the relation found by
Shimasaku (1997) for 40 clusters studied in Jones \& Forman
(1984) and Arnaud {\it et al.} (1992).

\section{Discussion}

For the SZ observational data in Table 1, 
we derived $\Delta T_{\rm SZ} \propto
L^{0.61 \pm 0.18}$.
The  slope is consistent with the numerically simulated values for
$\alpha$ ranging from 5 to 3, but slightly inconsistent with a $L-T$
relation of the form $L \propto T^2$, which is expected for clusters under the
self-similar model. The current
observational data suggest that $L \propto T^{2.6}$ (Markevitch 1998) to
$T^{2.8} $(Arnaud \& Evrard 1998),
which is consistent with the present estimate of the 
$\Delta T_{\rm SZ}-L$ slope. 

The $M_{\rm gas}-L$ relation, as derived based
on the observed mass data for a large sample of
clusters in White et al. (1997) suggests
a slope of $0.66 \pm 0.06$ to the $\Delta T_{\rm SZ}-L$ relation,
which is again consistent with the observed value.
Since the SZ temperature decrement is a direct estimate on
the cluster gas mass, and assuming that the measured gas
mass is essentially similar to the gas mass the SZ experiments
would observed based on the geometry, then $\Delta T_{\rm SZ}
\propto L^{0.66 \pm 0.06}$. However, effects due to cluster
projections and other associated systematic errors (see, below) may produce
a slightly different
observed relation that 
one expected based on these simple arguments.
We note that the observed $M_{\rm gas}-L$ relation is based
on the gas masses within inner 0.5 MPc from the cluster centers.
Depending on the redshift and the instrumental properties (beam size,
resolution etc), individual SZ experiments may be sensitive
to a different radius from the cluster center. The
$\Delta T_{\rm SZ}$ values are derived by modeling the observed flux
over a larger radius than 0.5 Mpc. Thus, comparsion of the two relations
may be slightly problematic, however, since the total gas mass
out to a larger radius would scale in a self-similar manner, we
do not expect the slope to be affected. However, comparsion of the
normalizations are not currently possible. Also, we have used
two different estimates of luminosities -- luminosities in 2 to 10
keV for $\Delta T_{\rm SZ}-L$ relation and bolometric
luminosities for $M_{\rm gas}-L$ relation ---  but
here again, we don't expect the slope to have been affected
systematically by these differences.

Other than statistical uncertainties associated with the measurement
of SZ decrements and  X-ray luminosities, 
it is
likely that there are additional effects contributing to the observed scatter
in the $\Delta T_{\rm SZ}-L$ relation.
Such contributions may come from projection effects of galaxy clusters,
which are mostly modeled using spherical geometries,
variations in gas mass fraction from one cluster to another, and effects due
to gas clumping and nonisothermality. 

For example, if a cluster
is prolate and elongated along the line of sight, then the observed
SZ temperature decrement would be larger than what is expected based
on the X-ray properties of that cluster (see, Cooray 1998). In such a
case, one may also find a substantially lower value for the Hubble
constant. Since we are considering the
galaxy cluster sample in a statistical manner, the derived 
relations such as $\Delta T_{\rm SZ}-L$,
should be unbiased of effects that may arise due to cluster
projections. However, this requires that the clusters in the sample
are distributed randomly, so that the whole sample is unbiased. 
This may not likely to be true with
the current sample, as most of the current SZ targets are 
clusters which are likely to have been previously studied due to
certain properties. Such
clusters, which may have been selected due to high X-ray luminosities
or gravitational lensing effects, are likely to be prolate  clusters
elongated along the line of sight
such that the observed decrement may be slightly higher
than the value expected for the luminosity of that cluster. 
This is more likely to be the case with strong gravitational lensing
clusters; the fact that both A1689 and A2163 have similar temperature 
decrements but different X-ray luminosities clearly suggests this possibility. 
Also, high luminosity clusters are likely to be strong
cooling flows, but the luminosity values used here, except for Cl0016,
 have been corrected
for such cooling flow contributions. For the present sample, if biased
effects exist either at the 
low or high end of the luminosities, then the derived relation
may have been affected.

From the present SZ sample, A2218, A773, and Cl0016 
have been observed by multiple
observational programs. The difference between these
separate measurements are of the order 15\% to 25\%. Other than
physical systematics effects discussed above, it is likely that
the derived relation has an additional  intrinsic dispersion 
as high as 25\%. Such differences are
likely to arise when clusters are modeled using different parameters
and that the observational effects, such as the beam dilution produced by the
instrument, are not properly taken into accounted. 
It is likely that the $\Delta T_{\rm SZ}-L$ relation will be properly
studied using carefully selected sample of galaxy clusters that are
planned to be observed with ground-based interferometers, 
and space-based observatories such as {\it PLANCK}.

The current cluster sample used to derive the $\Delta T_{\rm SZ}-L$
relation
ranges in redshift from 0.02 to 0.54, with only one cluster beyond
a redshift of 0.2. Therefore, we are unable to study any
redshift evolution of this relation. However, we note that when using this
relation to study clusters at high redshifts, it may be 
important to consider the possible evolutionary effects. 
However, if such effects
exist in the $\Delta T_{\rm SZ} - L$ relation,
then it would be primarily due to the evolution of the X-ray luminosity
function, but other effects, such as a systematic change in the
 cluster baryonic gas mass fraction with redshift,
can also produce deviations in the $\Delta T_{\rm SZ}-L$ relation. 

In order to test the accuracy of the derived relation between SZ
effect and X-ray luminosity of galaxy clusters, we now consider
additional observable cluster properties.
Since the X-ray luminosity and the gas mass
of a cluster is a measurement of the baryonic content,
the presented $\Delta T_{\rm SZ} - L$ is primarily a probe of the
baryonic mass distribution.
When the $\Delta T_{\rm SZ} - L$ relation is
combined with scaling relations such as
$L-T_{\rm e}$ and $M_{\rm tot} - T_{\rm e}$ (Hjorth, Oukbir \& Kampen 1998), 
one can constrain both the baryonic and
dark matter distribution separately from each other. In Fig.\ 3, we
show the observed temperature decrements as a function of the cluster
electron temperatures
in table 1. Even though the involved uncertainties are high, we find
that the $\Delta T_{\rm SZ}-T_{\rm e}$ is:
\begin{equation}
\Delta T_{\rm SZ} = -(0.56 \pm 0.51) \times 10^{-3} \left(\frac{T_{\rm
e}}{{\rm keV}}\right)^{2.35 \pm 0.85}\, {\rm mk}.
\end{equation}
The slope of this relationship is again consistent with simple
scaling-law arguments. 

\begin{figure}
\vspace{5cm}
\caption{The observed $\Delta T_{\rm sz}$ vs. $T_{\rm e}$ relation,
with the best fit shown as a solid line. The SZ cluster sample is represented
with same symbols as in Fig.~1.}
\end{figure}

The luminosity-gas mass relation observed from the X-ray
measurements in Fig.\ 2, suggests $\Delta T_{\rm SZ} \propto
L^{0.64 \pm 0.06}$, which is in good agreement with 
the relation derived from the SZ observational data.
As suggested earlier, this relation is more likely to be accurate
since the integrated pressure along the line of sight is simply
proportional to the gas mass within the line of sight, which is the gas mass
within the cylinder defined by geometry. A direct probe of the total cluster mass along the
line of sight is the mass derived based on gravitational lensing
observations. Smail et al. (1997) studied
lensing properties of a sample of galaxy clusters observed with the
{\it Hubble Space Telescope} and measured both weak and strong lensing properties of
the clusters. They derived a relationship between cluster luminosity, 
$L$, and mean shear, $<\gamma>$, of the form:
\begin{equation}
<\gamma> = (0.074 \pm 0.017) \times \left(\frac{L}{10^{44}\,
h_{50}^{-2}\, {\rm ergs\, s^{-1}}}\right)^{0.58 \pm 0.23}.
\end{equation}
Here, $\gamma$ is a measurement of the average tangential shear
strength of galaxy clusters, and is directly proportional to the
cluster total mass responsible for gravitationally lensing the
background galaxies towards the foreground cluster.
Thus, total mass can be written as $M_{\rm tot} \propto L^{0.58 \pm 0.23}$.
Since $M_{\rm gas} \propto \Delta T_{\rm SZ} \propto L^{0.66 \pm 0.06}$,
we find that the ratio, $f_{\rm gas} \equiv M_{\rm gas}/M_{\rm tot}$,
is $\propto L^{0.08 \pm 0.24}$. This ratio measures the cluster gas
mass fraction, which has been used in literature to constrain the
total mass density of the universe based on baryonic-mass density
as derived based on nucleosynthesis arguments (see, e.g., Evrard 1997). 
It is likely that this ratio is independent of the cluster luminosity,
suggesting that the gas mass fraction within clusters is constant
from one cluster to another. Recently, Arnaud \& Evrard (1998) studied
the changes in cluster gas mass
fraction from one cluster to another and suggested that these changes
are likely be due to heating processes within clusters, such as due to
winds. If such effects exist, then the $\Delta T_{\rm SZ}-L$
relation would also be affected contributing to the observed
dispersion.

We use the X-ray luminosity function (XLF) of clusters of galaxies
from the {\it ROSAT} Brightest Cluster Sample (Ebeling et al. 1997), which is an X-ray
selected, flux limited sample of 172 clusters compiled from the {\it
ROSAT} All-Sky Survey data, to study the local cluster contribution to
arc-minute scake
cosmic microwave background (CMB) anisotropies. 
In order to compare with the current limits of the Compton
$y$-parameter based on {\it FIRAS} observations, we convert our
$\Delta T_{\rm SZ}-L$ relation to a $y-L$ relation using:
\begin{equation}
y(L) = -\frac{1}{2} \frac{\Delta T_{\rm SZ} (L)}{T_{\rm CMB}}.
\end{equation}
The XLF is represented by:
\begin{equation}
\phi(L)dL = A L^{-\alpha} \exp{\left(-\frac{L}{L_\star}\right)}dL,
\end{equation}
where $A=(1.59 \pm 0.36) \times 10^{-7}\, h_{50}^3\, {\rm
Mpc^{-3}} \times (10^{44} h_{50}^{-2} {\rm ergs\,
s^{-1}})^{\alpha-1}$, $L_{\star} = (8.46 \pm 2.35) \times 10^{44}\,
h_{50}^{-2}\, {\rm ergs\, s^{-1}}$ and $\alpha=1.25 \pm 0.20$.
Here, $L$ is the luminosity in the 2 to 10 keV band.
Using the luminosity function and the $y(L)$ relation, we can write
the average $<y>$ parameter due to galaxy clusters as:
\begin{equation}
<y> = \int_{L_1}^{L_2} y(L) \phi(L) dV dL
\end{equation}
where $L_1$ and $L_2$ are the lower and upper limits of XLF,
respectively. In order to describe the cluster volume $dV$ we adopt a
$\beta$-model for the cluster gas distribution with a core radius
$R_c$. We assume a $\beta$ of 2/3 to describe all clusters. We allow
for the total cluster scale radius $R$ to vary with luminosity of the
form, $R \propto L^\kappa$, where $\kappa$ is a free-parameter. 
According to Mohr \& Evrard (1997), the X-ray size scales
with temperature as $R \propto T^{0.93 \pm 0.11}$. With the $L-T$
relation, the X-ray size is expected to scale with luminosity as
$R \propto L^{0.2}$ to $L^{0.5}$. Using the tabulated data in White
et al. (1997), we obtained a relation of the form, 
\begin{equation}
R_{\rm c}(L) = 0.46 \left(\frac{L}{10^{44}\, h_{50}^{-2}\, {\rm ergs\,
s^{-1}}}\right)^{0.33 \pm 0.15}\, {\rm Mpc},
\end{equation} 
between the tabulated core radii for clusters, $R_{\rm c}$, 
and their luminosities, $L$. 
We note that the core-radii from White et al. (1997) do not
necessarily represent the true underlying sacle-size
of the X-ray emission. In their analysis, each cluster
core-radius was
treated as a  parameter which was varied to obtain
a flat temperature profile, under their assumption for the
form of the gravitational potential. For the purpose of the
present calculation, however, we are only interested in core-radii 
as a representation for the relative distribution of
size-scales. Since we also 
vary most of the parameters, such as the slope in the $R-L$ relation,
and the fact that our final results our presented as a ratio between the
cluster radius and the core-radius, the use of a $R-L$ relation  from data in
White et al. (1997) should not affect the final conclusions.
For the rest of the paper we assume that $\kappa$ ranges from 0 to 1.
In order to test the accuracy of derived $\Delta T_{\rm SZ}-L$
relation, we also vary its slope $\gamma$, between 0 and 2.5. 

In Fig.~4 we show the derived $<y>$  values as a function of
$R/R_c$. It is likely that cluster sizes, in general, are of the order
15 to 25 core radii. Each of the individual plots correspond to the
different $R-L$ relation at steps of 0.25 between 0 and 1, while each
of the 
curves represent the different $\gamma$ values at steps of 0.5 between
0 and 2.5.  For our preferred values of $\kappa$ and $\gamma$,
$\kappa \sim 0.25$ to 0.5 and $\gamma \sim 0.5$ to 1.0,
we find that the cluster contribution to $y$-parameter through the SZ
effect is at least two orders of magnitude less than the current
upper limit based on {\it FIRAS} observations of $2.5 \times 10^{-5}$
(Mather et al. 1994). If all clusters have a
constant size, independent of the X-ray luminosity, then the $<y>$ can be as
high as $10^{-6}$, but this possibility 
is excluded by the observational data, which suggests that cluster
size varies with the luminosity. In general, 
we find that the cluster contribution to $y$-parameter
is $\sim 2 \times 10^{-7}$, which is consistent with previous
estimates (see, e.g., Ceballos \& Barcons 1994). 
The XLF used here has been calculated for
clusters out to redshifts of 0.3, so that 
the derived results may be valid to clusters out
to the same redshift. However, evidence for no evolution in the XLF out to
redshifts of 0.8 has recently 
appeared on literature (Rosati et al. 1998), and thus, our results may
be valid to a much higher redshift. As studied in Barbosa et al. (1996),
in a low $\Omega_m$ universe, galaxy cluster contribution to Compton
$y$ parameter may be as high as $10^{-5}$, with most of the
contribution coming from clusters at $z > 1$.

\begin{figure}
\vspace{5cm}
\caption{$<y>$ as a function of $R/R_c$. The different plots
correspond to the different dependences between cluster size scale
and the luminosity, while individual curves within plots represent
different dependences between SZ temperature decrement and cluster
luminosity. For the preferred values of $\kappa$ and $\gamma$ (see
text), the cluster contribution to arcminute scale anisotropies
are few times $10^{-7}$.}
\end{figure}

Other than using the $\Delta T_{\rm SZ}-L$ relation as a probe of
cluster physics, the observed $\Delta T_{\rm SZ}-L$ relation can also 
be used in the study of high redshift  clusters, where SZ temperature
decrements have been observed but no X-ray emission have been detected. For example, 
given that we now know the $\Delta T_{\rm SZ}-L$ relation, 
it is reasonably possible to constrain the expected X-ray luminosity
of such a detection,  and then, perhaps, based on the observed X-ray flux threshold, 
put a lower limit on the redshift. 
Detection of such high redshift clusters has strong implications
for cosmological world models, and
based on the expected redshift and the mass of such clusters,
one can constrain cosmological parameters with high accuracy (e.g,
Bartlett, Blanchard \& Barbosa 1998). The $\Delta T_{\rm SZ}-L$ and 
$M_{\rm gas}-L$
allows one to directly relate the observed
temperature decrements to gas mass, and under the assumption
of a constant baryonic fraction, to relate to the total mass of the
cluster. It should then be possible to apply the 
Press-Schechter (PS) formalism to number density of
such cluster masses, and constrain the cosmological parameters,
 with little dependence on the X-ray observations. 
However, to perform such an analysis one requires data from a complete
sample of galaxy clusters, and such SZ samples are currently not
available; observations of a luminosity-selected unbiased
sample of galaxy clusters would be useful in the future to constrain 
the cosmological parameters, using scaling relations such as the one 
presented here.

\section{Conclusions}
Based on observations of the Sunyaev-Zeldovich effect in galaxy clusters
and the X-ray luminosity, we have derived a 
relation between the two observables. 
We have studied this relation in terms of other cluster properties,
and have found it to agree with the $L-T$, $M_{\rm gas}-L$ and
$M_{\rm tot}-L$ relations. Using the observed $\Delta T_{\rm SZ}-L$
relation and the X-ray luminosity function for galaxy clusters, we
have derived the local cluster contribution to the Compton
$y$-parameter through the SZ effect. These values are at least two
orders of magnitudes lower than the current upper limit based on {\it
FIRAS} observations. Given the planned observations of SZ effect in
large sample of galaxy clusters, such as with {\it PLANCK} and other
ground based interferometers, it is likely that a relation such as
$\Delta T_{\rm SZ}-L$ would be useful both in making predictions and
deriving important cosmological parameters. We leave such studied to
be carried out in future, using complete and unbiased samples of
galaxy clusters.

\section*{Acknowledgements}
I would like to thank David A. White, the referee, for constructive
comments on the manuscript, John Carlstrom and
Bill Holzapfel for a helpful suggestions on an early draft of the
paper and Dan Reichart for useful discussions.

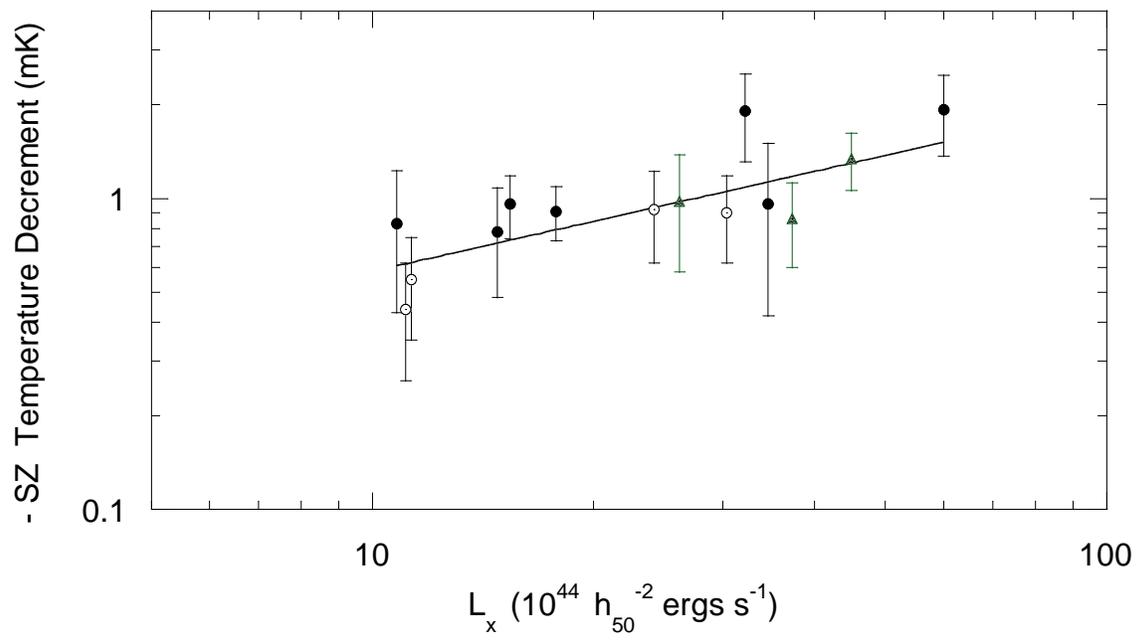
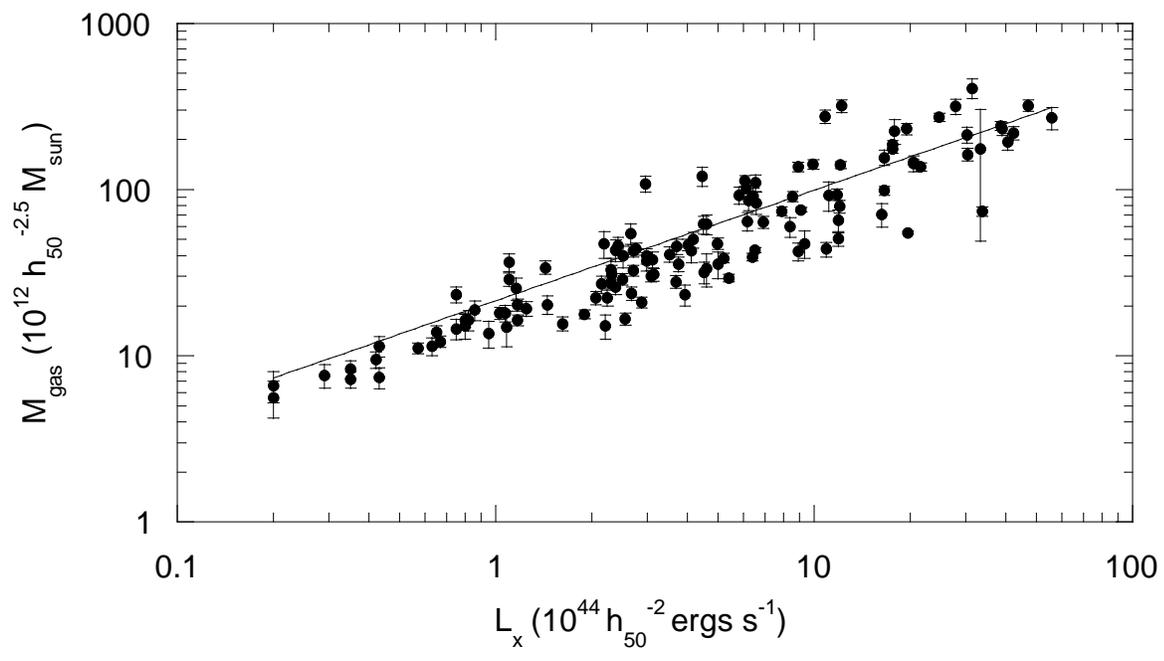
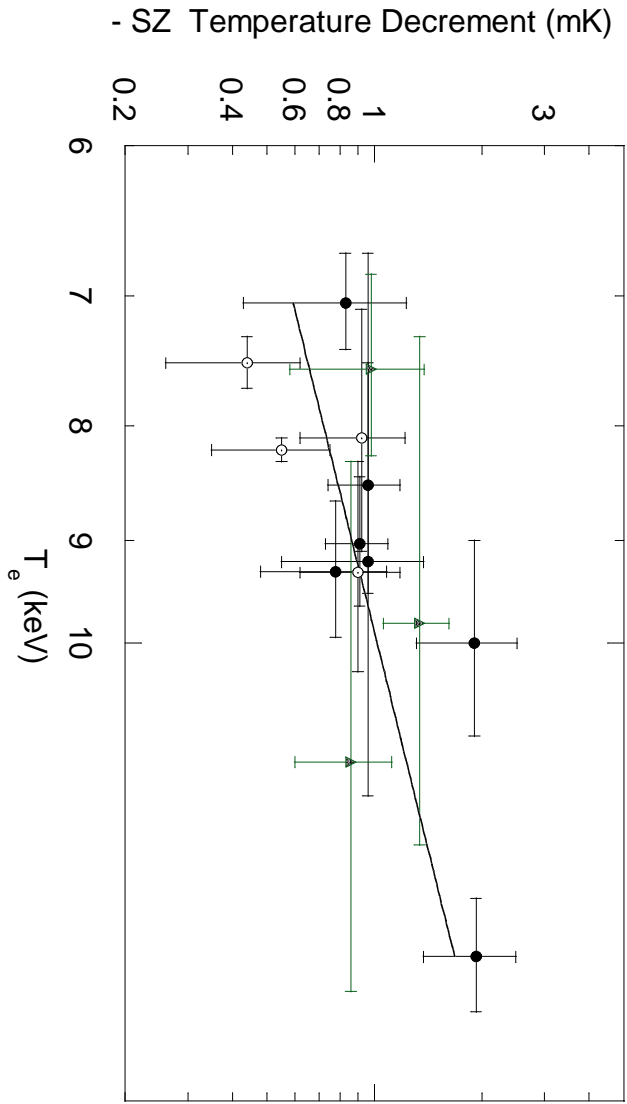
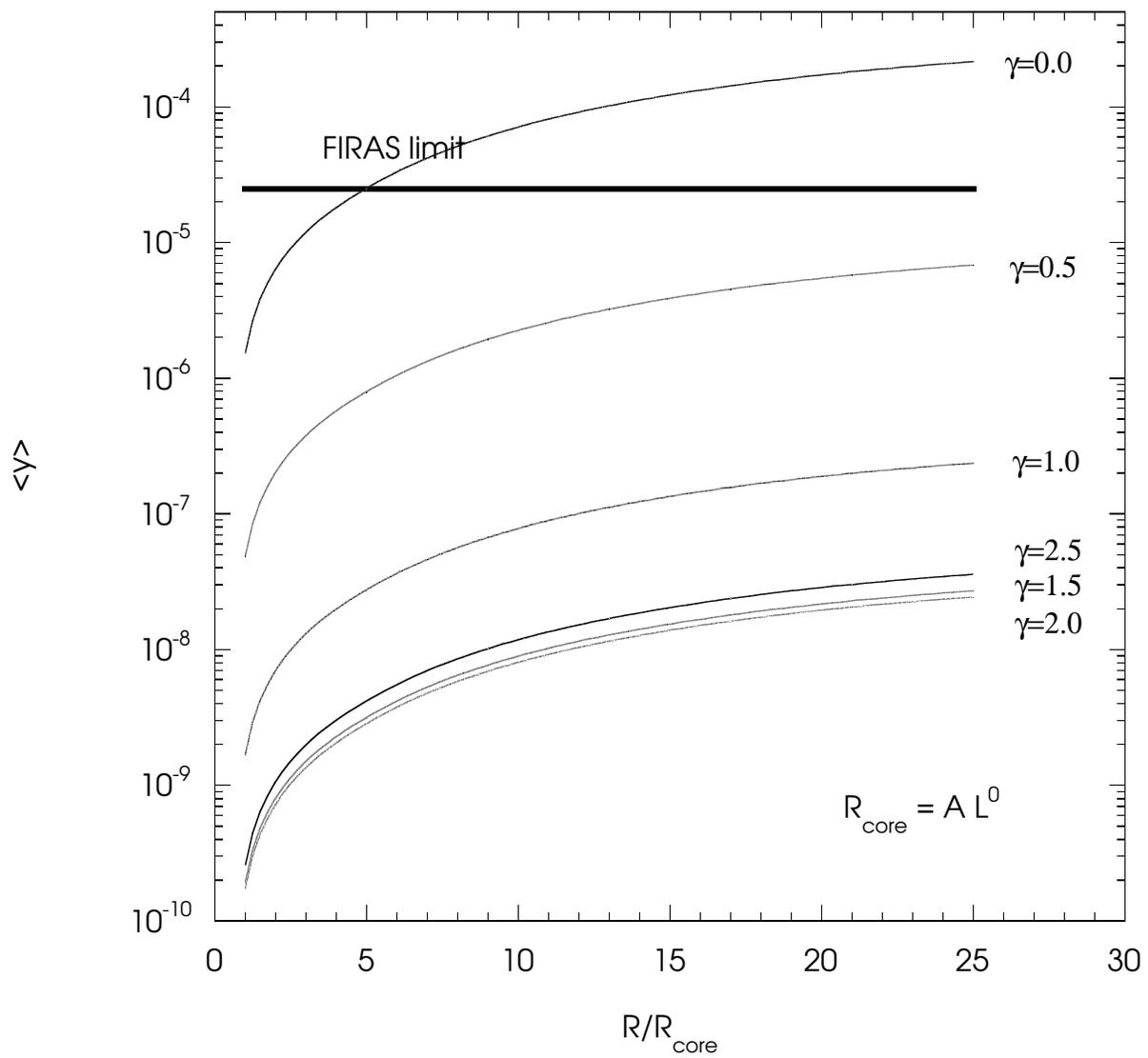
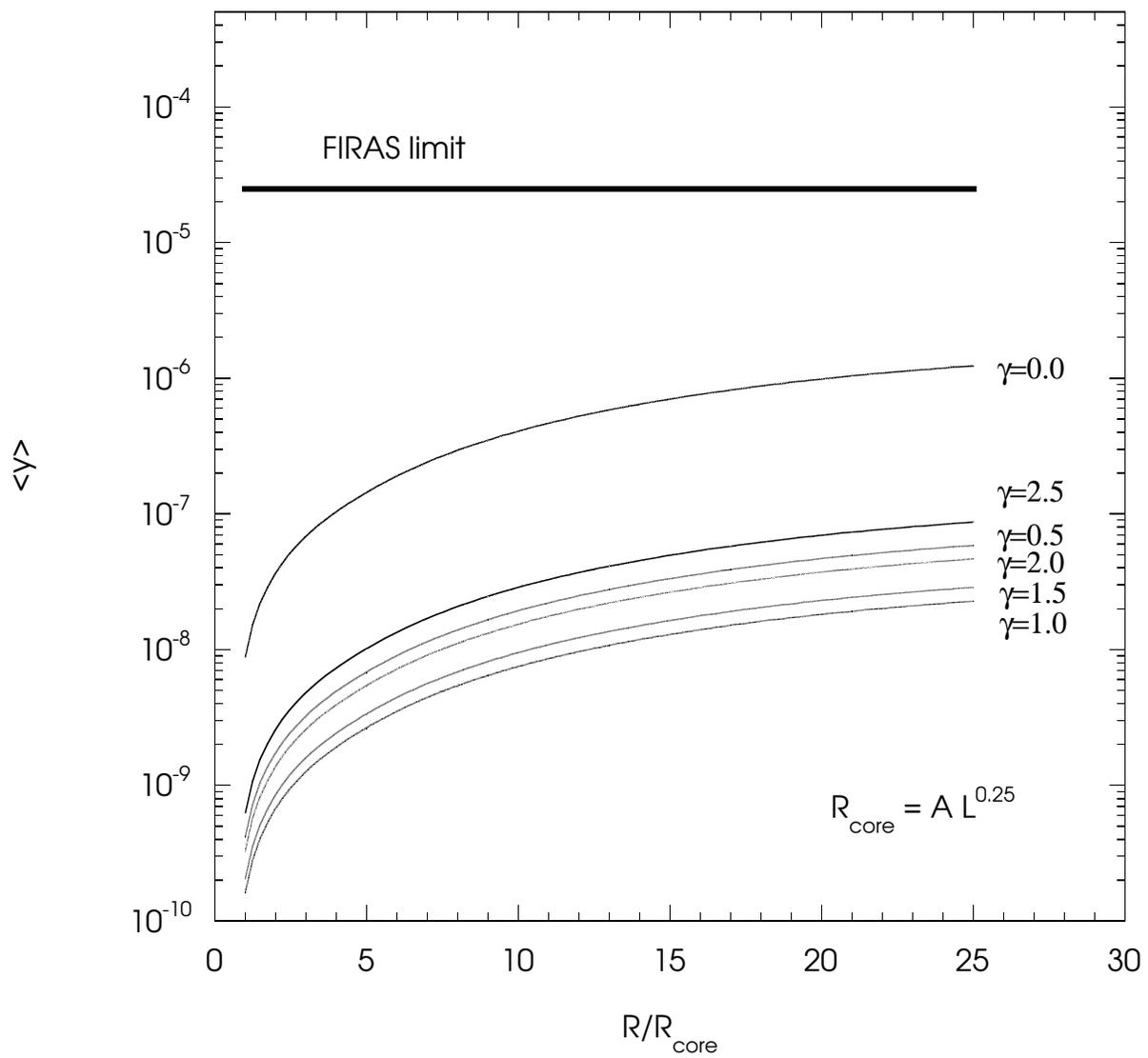
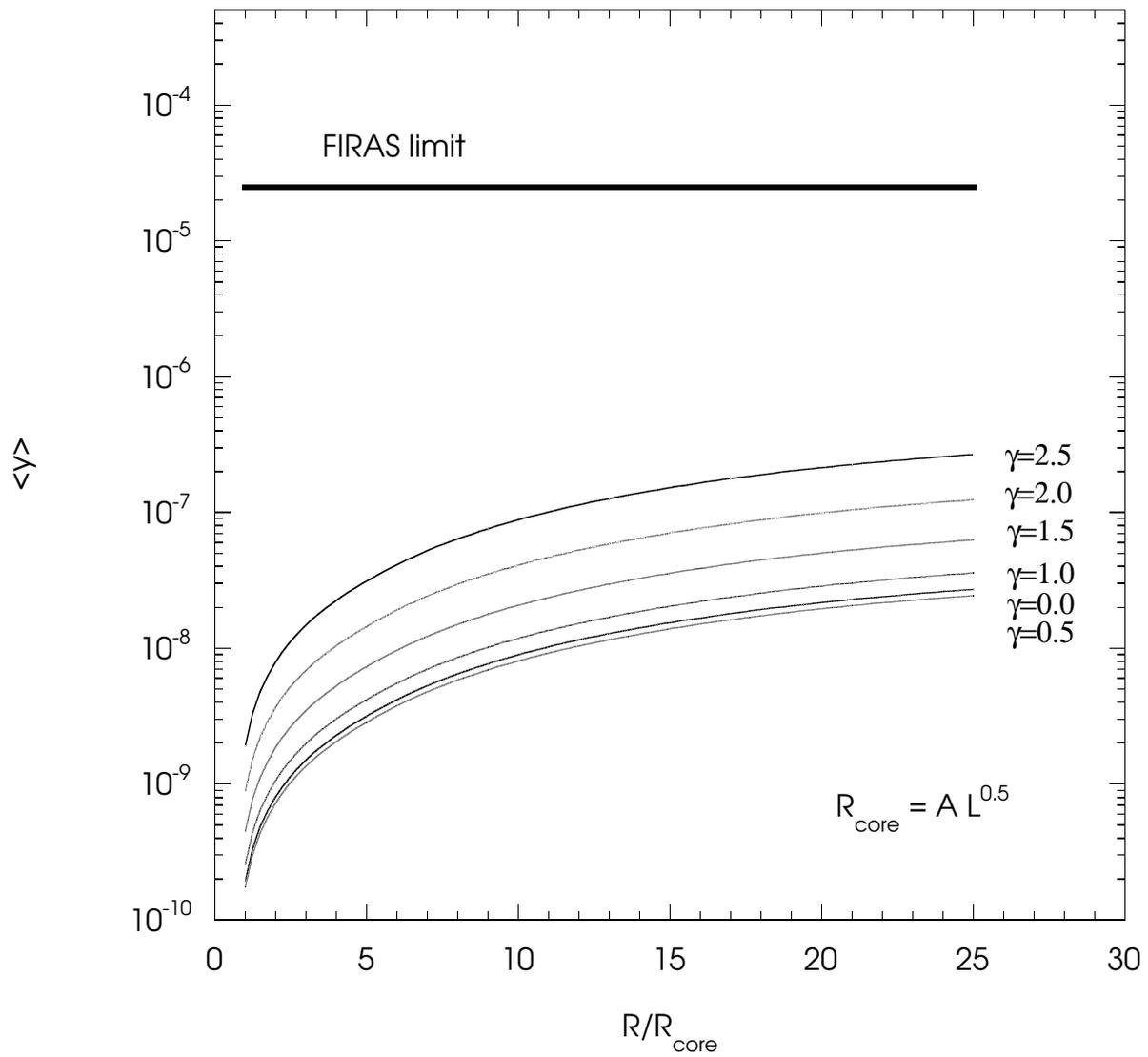
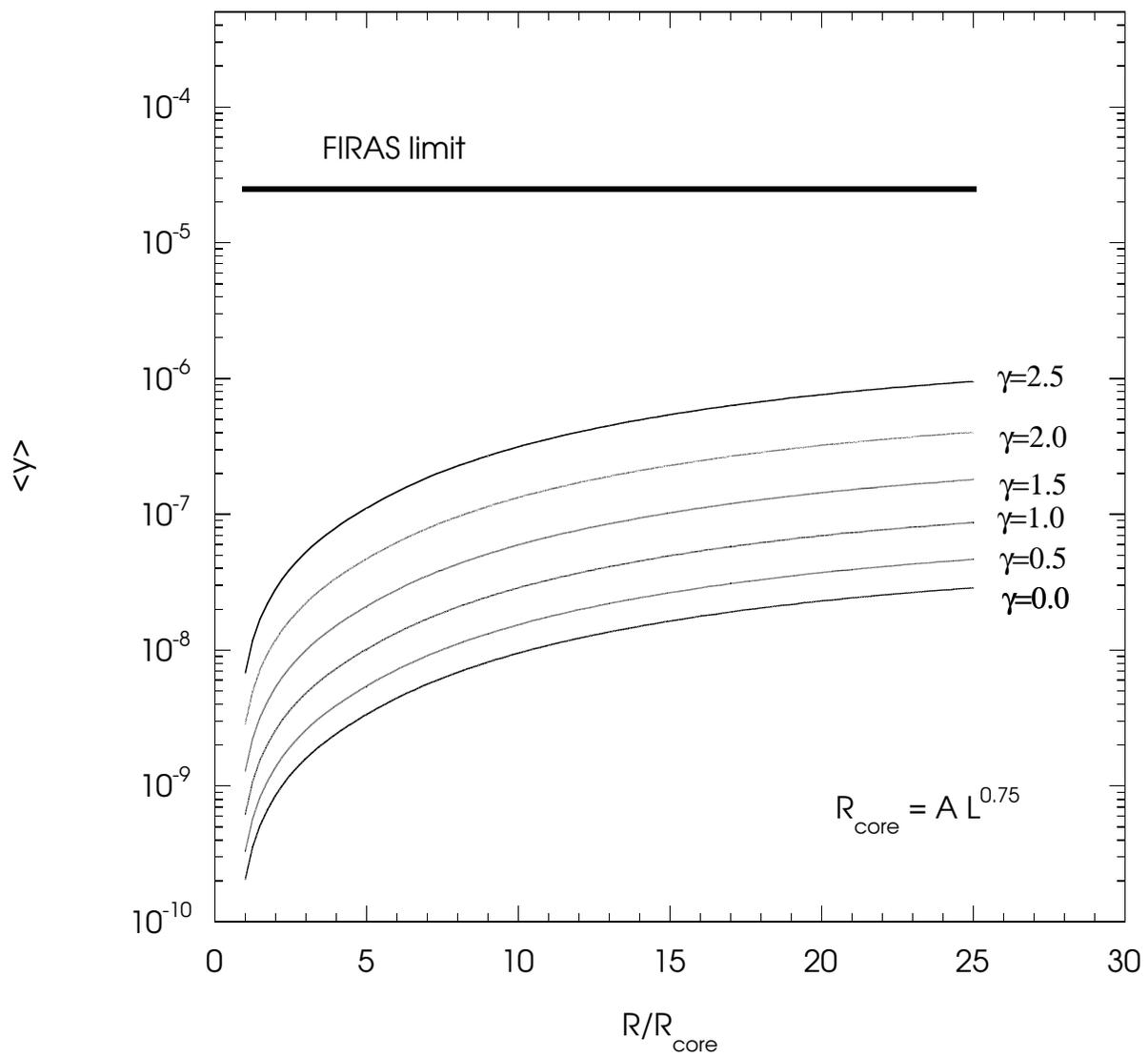
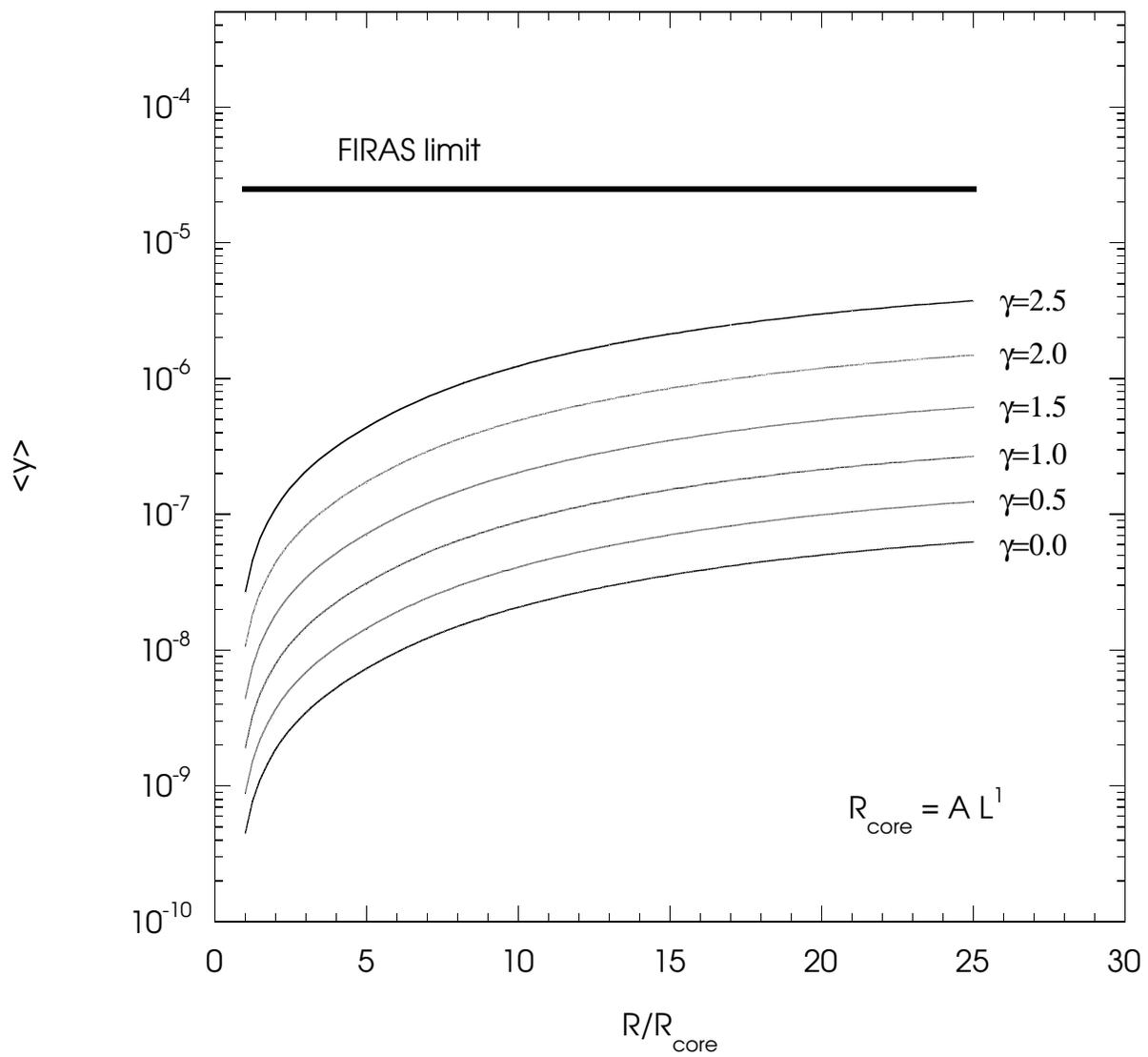
\end{document}